\documentclass{article}
\usepackage{psfig}
\usepackage{emulateapj}

\lefthead{Nevalainen J., Markevitch M and  Forman W.}
\righthead{The cluster M-T relation from temperature profiles observed with ASCA and ROSAT}

\begin{document}

\submitted{Accepted to ApJ}

\title{The cluster $M -T$ relation from temperature profiles observed with ASCA and ROSAT}

\author{J. Nevalainen\altaffilmark{1}, M. Markevitch\altaffilmark{2} and W. Forman}
\affil{Harvard Smithsonian Center for Astrophysics, Cambridge, USA}

\altaffiltext{1}{Observatory, University of Helsinki, Finland}
\altaffiltext{2}{Space Research Institute, Russian Acad. of Sci.}

\begin{abstract}

We calibrate the galaxy cluster mass - temperature relation using the temperature profiles of intracluster gas observed with ASCA (for hot clusters) and ROSAT (for cool groups). Our sample consists of 
apparently relaxed clusters for which the total masses are derived assuming hydrostatic equilibrium.  The sample provides data on cluster X-ray emission-weighted cooling flow-corrected temperatures and
total masses up to $r_{1000}$. The resulting M-T scaling in the 1-10 keV temperature range is
$M_{1000} = 1.23 \pm 0.20 \ 10^{15} \ h_{50}^{-1} \ M_{\odot}
\left({{\langle T \rangle_{z = 0}} \over {\rm 10 \ keV}}\right)^{1.79 \pm 0.14}$
with 90\% confidence errors, or significantly (99.99\% confidence) steeper than the self-similar relation $M \propto T^{3/2}$. For any given temperature, our measured mass values are significantly smaller 
compared to the simulation results of Evrard et al. (1996) that are frequently used for mass-temperature scaling. The higher-temperature subsample (kT $\ge$ 4 keV) is consistent with M $\propto$ T$^{3/2}$, 
allowing the possibility that the self-similar scaling breaks down at low temperatures, perhaps due to heating by supernovae that is more important for low-temperature groups and galaxies as suggested by 
earlier works.

\end{abstract}


\section{Introduction}
Galaxy clusters are the largest gravitationally bound objects in the universe, and thereby provide information on cosmic structure formation. The mass distribution of virialized objects can be predicted for
different cosmologies and different initial density fluctuation spectra. By comparing such predictions to the observed cluster mass function, one can constrain cosmological parameters. Cosmological 
parameters most strongly affect mass function predictions for large masses that correspond to the cluster scale, and therefore clusters are of special importance. 

Accurate measurements of the cluster total mass, dominated by dark matter, are challenging at present and possible only for a limited number of clusters. For this reason there is currently insufficient data 
for a direct derivation of the mass function. A more practical way of determining the mass function is to observe the distribution of readily available average cluster gas temperatures and to convert this to
a mass function, taking advantage of the tight mass - temperature correlation predicted by hydrodynamic cluster formation simulations (e.g. Evrard, Metzler \& Navarro 1996). The mass and temperature are 
predicted to scale as $M \propto T^{3/2}$. Although different simulations and observations are in general qualitative agreement, there are significant disagreements on the details of the gas temperature 
profiles (Frenk et al. 1999, Markevitch et al. 1998). Therefore, this relation needs observational confirmation and calibration.

In addition to being useful for providing a link between cosmological predictions and observations, the M-T relation is also interesting in itself, because any deviations from the predicted self-similar 
scaling of $M \propto T^{3/2}$ would indicate that additional physical processes are at play than gravity alone. The detailed behaviour of the M-T relation provides information about the process of cluster 
formation and energy input into the gas. One particularly important energy input source is preheating of the intergalactic gas by early supernova driven galactic winds (e.g. David, Forman \& Jones (1991), 
Evrard \& Henry (1991),  Kaiser (1991), Loewenstein \& Muskotzky (1996)).

Steps toward calibrating the M-T relation with observations have been made by measuring cluster masses using gravitational lensing (Hjorth, Oukbir \& van Kampen 1998) and the hydrostatic equilibrium 
approach assuming isothermal gas (\cite{neu}). Horner, Mushotzky \& Scharf (1999) derived M-T relation using several independent methods: virial theorem for cluster galaxies and hydrostatic equilibrium
applied to the X-ray emitting gas, both assuming isothermality and using published mass values derived from measured 
temperature profiles.

As in the latter works, in this paper, we derive a cluster mass - temperature relation under the assumption of hydrostatic equilibrium. We use the published total mass profiles and X-ray emission-weighted 
temperatures derived from temperature profiles of hot clusters measured with ASCA. For cooler groups, using the published ROSAT cluster temperature profiles, we compute the corresponding temperature values, 
and the total mass profiles in cases where masses are not published in sufficient detail. We incorporate several improvements over the work of Horner et al. (1999). Our ASCA cluster sample is homogeneous, 
the temperature profiles are all determined using the same method that accounts for the ASCA PSF (Markevitch et al. 1998) so that the resulting mass values and their errors are directly comparable. The 
Horner et al. (1999) sample contains mass profiles derived both with ASCA PSF correction (e.g. A2256 and A2029) as well as without it (e.g. A496 and A2199). For A496 and A2199 the PSF correction is not large
for the central pointing used in Horner et al. (1999), but we measure the temperature profile to a larger radius using offset pointings where the PSF correction is significant. Another important difference 
is that Horner et al. (1999) extrapolate mass profiles to an overdensity radius $r_{200}$ assuming $\rho_{dark} \propto r^{-2.4}$, whereas we use the measured mass profiles up to $r_{1000}$. At this radius 
relaxed clusters are unlikely to experience significant residual turbulence that would violate the hydrostatic equilibrium assumption (e.g. Evrard et al. 1996). Our sample contains only clusters without 
signs of disturbance. Thirdly, we combine our ASCA sample with a low-temperature subsample of galaxy groups and galaxies, whose temperature profiles have been measured with ROSAT, in order to study the M-T 
relation over a temperature range of 1 - 10 keV.

We use $H_{0} \equiv 50 \, h_{50} \, {\rm km} \, {\rm s}^{-1} \, {\rm Mpc}^{-1}$, $\Omega = 1$ and report 90\% confidence 
intervals throughout the paper, except where stated otherwise.

\section{DATA}

\subsection{The sample}
The sample consists of 9 clusters, groups and galaxies with published (see Section 2.3) relatively accurate spatially resolved temperature profiles measured with ASCA (the 6 hotter ones) and ROSAT (the 3 
cooler ones) up to radii of overdensity of 1000 - 500, comparable to the virial radii. Hereafter we use the usual notation $r_{i}$ and $M_{i}$ where $r_{i}$ is a radius of overdensity (the mean interior 
density with respect to the critical density) of $i$, and $M_{i}$ is the mass within that radius. For our sample, we required the objects to be apparently relaxed, with no substructure or deviation from 
azimuthal symmetry in ROSAT PSPC images, or in the ASCA temperature maps in Markevitch et al. (1998). Hence, the effects of bulk motions, and the deviation from hydrostatic equilibrium are minimized.
The requirement of apparent relaxation limits severely the number of suitable cluster candidates for our analysis. 

Most relaxed clusters have strong cooling flows which complicate the ASCA spatially resolved analysis. Since the ASCA PSF has a half-power diameter comparable to the angular size of a typical cooling flow in
a nearby cluster, the cooling flow parameters can not be constrained adequately. However, when modeling the temperature structure of the cluster, the (uncertain) cooling flow model must be included as a 
component. The wide energy-dependent PSF scattering of the photons from a strong cooling flow therefore increase the uncertainty of the non-cooling-flow model temperatures even at a large radius. Therefore 
we mostly selected relaxed clusters with only moderate cooling flows (and therefore accurate temperature profiles), which further limited the number of suitable sample clusters. 

We seek relaxed systems which span the 1-10 keV temperature range. The clusters with the lowest temperatures are also the faintest and accurate spectroscopy is possible only for the nearest ones. The obvious 
problem is that in nearby clusters the virial radius corresponds to a large angular distance, usually beyond the ASCA and ROSAT field of view. This furthermore limits the number of available objects at low 
temperatures (kT $<$ 4 keV). There are several nearby cool groups and galaxies, for which the analysis with the needed accuracy is feasible with ROSAT, with a slight extrapolation of the mass profiles 
(see 2.4.).

Our search in the archives and literature resulted in the following sample of objects suitable for our analysis: A496, A2199, A401, A3571, A2256 and A2029 studied with ASCA, and NGC5044, NGC507 and HCG62 with
ROSAT.

\subsection{The mass errors}
Total masses are determined assuming hydrostatic equilibrium using the measured gas temperature and density profiles. For some clusters the formal uncertainty of the resulting mass values is small. However, 
hydrodynamic simulations show that the systematic uncertainties inherent in the hydrostatic mass determination method, such as deviations from spherical symmetry or hydrostatic equilibrium due to incomplete 
thermalization of the gas, will lead to about a 10 - 30\% uncertainty in the calculated total masses (e.g. Evrard et al. 1996; Schindler 1996; Roettiger, Burns \& Loken 1996). Since these simulations include
also mergers, whereas our sample contains only relaxed clusters, our uncertainties would be towards the low end of the above interval. However, to be conservative, if any formal mass error from the 
literature is smaller than 20\% of the mass value we use a 20\% uncertainty instead. 
The errors on the emission-weighted temperatures are much smaller than the mass errors. For example, at the most interesting radius $r_{1000}$ (see below) most of the relative temperature errors are smaller 
than 0.2 times the relative $M_{1000}$ errors wherefore we ignore the temperature errors while deriving the M-T relation.
 
\subsection{The ASCA subsample}
\label{asca}
For the hotter ($>$ 4 keV) clusters (A496, A2199, A401, A3571, A2256, A2029) we use the emission weighted cooling-flow corrected ASCA temperatures from Markevitch et al. (1998). For these clusters we use the 
published total mass profiles obtained using observed ASCA temperature profiles and ROSAT surface brightness profiles.  The masses of A496 and A2199 (\cite{a496_a2199}) have been derived by modeling the 
temperature profile with a polytropic form and the mass for A2029 (\cite{sar2029}) has been derived by modeling a temperature profile as a linear function of the radius. For A401 (Nevalainen, Markevitch \& 
Forman 1999a), A3571 (Nevalainen, Markevitch \& Forman 1999b) and A2256 (Markevitch \& Vikhlinin 1997) the masses have been obtained modelling the dark matter density with different functional forms (see 
above papers for the details), computing the corresponding temperature profiles and fitting the dark matter distribution parameters. For all these clusters the temperature profile is determined relatively 
accurately out to $r_{500}$. All these clusters exhibit a temperature decline with radius, which results in smaller total mass values at our radii of interest, compared to the usual isothermal analysis.
The ``universal'' NFW dark matter profile (Navarro, Frenk \& White  1997) provides a good description of these profiles over a large range of radii.

\subsection{The ROSAT subsample}
For the cooler objects, we used the published ROSAT temperature and surface brightness profiles to compute the emission weighed mean temperatures outside the cooling flow regions, to be directly comparable to
the cooling flow-corrected ASCA temperatures for hotter clusters. We wish to do all the analysis at $r_{500}$, but the ROSAT PSPC data do not extend to such large radii. We extrapolate the mass profiles 
beyond the radii of measured temperatures (except in the case of NGC507, where the temperature profile extends sufficiently far), but not beyond the radii to which the cluster surface brightness is 
significantly detected. For the group HCG62 and the galaxy NGC507 this allows us to reach $r_{1000}$, but for the NGC5044 group we reach only $r_{1500}$ (see Figure 1). While for the hotter clusters we use 
the published mass and temperature values, ROSAT data need some additional analysis, as described in detail below.
  
\subsubsection{Group HCG62}
\label{HCG62}
Using the ROSAT PSPC temperature profile from Ponman \& Bertram (1993) we obtain the X-ray emission-weighted temperature $\langle T \rangle$ = 1.16 $\pm$ 0.08 keV beyond the cooling radius of $r = 2.4'$.
For comparison, a published single component temperature of ASCA data for HCG62 gives a smaller value of $\langle T \rangle$ = 1.05 $\pm$ 0.02 keV (\cite{fuka}). However, in the ASCA analysis the central 
0.15$h_{50}^{-1}$ Mpc was excluded, which corresponds to 6.4$'$ for HCG62. Thus the ASCA value integrates over the outer parts of HCG62. In that region, the declining ROSAT temperature profile is consistent 
with the above ASCA value. Therefore, we use the value derived from the profile of Ponman \& Bertram (1993).

To derive the mass profile, we fitted the above ROSAT temperature profile with a polytropic form ($ T(r) \propto \rho_{gas}(r)^{\gamma -1}$), fixing the gas density using the results of the ROSAT surface 
brightness analysis (Ponman \& Bertram 1993), and found the best fit with $\gamma = 1.09^{+0.15}_{-0.14}$. For this fit, we excluded the cooling flow region $r < 2.4'$. The total mass for a polytropic
temperature profile inside a radius $r = x \ a_{x}$ is given by
\begin{equation}
M_{tot}(r) = 3.70 \times 10^{13} \ M_{\odot} \ T(r)  \
a_{x} \ {{3 \beta \gamma x^{2}} \over {1 + x^2}} {\mu \over 0.60},
\end{equation}
where $\mu$ is the mean molecular weight, $\beta$ and $a_{x}$ are the slope parameter and core radius of the gas density profile, T is expressed in keV and  $a_{x}$ in Mpc. We propagated the errors of 
$\gamma$ and T to the total mass.

\subsubsection{Galaxy NGC507}
Kim \& Fabbiano (1995) obtained a temperature profile for this galaxy with ROSAT PSPC. Excluding the data from the cooling flow area ($r < 2.0'$), we obtain a mean temperature of $\langle T \rangle$ = 1.09 
$\pm$ 0.03 keV. The corresponding value obtained by ASCA is significantly higher, $\langle T \rangle$= 1.26 $\pm$ 0.07 keV (\cite{fuka}). We are not strongly concerned about this disagreement, because the 
ROSAT and ASCA energy bands are very different and there may be differences in the average temperature due to, for example, nonisothermality and a possible nonthermal component. We assume that for low 
temperature systems ( $\langle T \rangle$ $\sim 1$ keV), ROSAT gives a more accurate value, and use the above ROSAT value in our analysis. We show below that if we use the higher ASCA temperature value 
instead, our conclusions will only be strengthened.

The total mass profile is directly taken from Kim \& Fabbiano (1995), who used a polytropic model for the temperature profile, as described in (\ref{HCG62}) above. The errors of temperature and the polytropic
index $\gamma$ are incorporated in the mass values.

\subsubsection{Group NGC5044}
Using the ROSAT temperature profile of David et al. (1994), we obtained the emission weighed temperature $\langle T \rangle$ $ = 1.09 \pm 0.03 $ keV, beyond the cooling radius $r = 4.0'$. ROSAT temperatures 
beyond 0.15$h_{50}^{-1}$ Mpc are consistent with the corresponding ASCA value ($\langle T \rangle$ = 1.07 $\pm$ 0.01 keV \cite{fuka}).

We use the analytical form for the best fit total mass given in David et al. (1994), obtained using a power-law model for the temperature profile. We propagate the errors of temperature and the exponent of 
the temperature profile to the total mass errors. Because the X-ray emission is not detected beyond $r_{1500}$ for this nearest group, we will not use this group in our analysis beyond that radius.  

\subsubsection{NFW mass profile}
The NFW model for the dark matter density profile describes well the mass distribution in hot ASCA clusters in our sample (see e.g. Markevitch \& Vikhlinin 1997, Sarazin et al. 1998, \cite{a496_a2199},
Nevalainen et al. 1999a, Nevalainen et al. 1999b). Furthermore, a similar result was found for Coma in an optical study of galaxies (Geller, Diaferio \& Kurtz 1999). Therefore, we compared the NFW model with
the derived mass profiles for the cool groups and galaxies HCG62, NGC507 and NGC5044. We fixed the gas density profile to ROSAT values and modeled the dark matter density distribution with an NFW profile 
selecting its parameters so that gas plus dark matter approximate the derived total mass profiles. For all three systems, the NFW profile provides a good description of the dark matter distribution over a
range of interesting radii (see Figure 1). Such an underlying universal dark matter profile is consistent with the observed similarity of gas density profiles (\cite{neu}, Vikhlinin, Forman \& Jones 1999) 
and temperature profiles (Markevitch et al. 1998) in different clusters, when scaled to physical radii.

\section{RESULTS}
For each object, using its measured total mass profile, we computed the overdensity, or the mean interior density in units of the critical density as a function of the radius: 
${M_{tot}(r)} / ( \frac{4}{3} \pi r^{3} \rho_{c})$, where $ \rho_{c} = 3 H_{0}^{2} (1 + z)^{3}/8 \pi G $. We then calculated the masses within several radii of fixed overdensity (see Table 1). 
To remove the different amounts of evolution in our sample temperatures due to different values of $z$, we scaled the observed mean temperatures to $z = 0$ according to the predicted scaling 
$T_{gas}(z) \propto (1 + z)$ for a given mass (e.g. Eke, Cole \& Frenk 1996). To be consistent with this, we should also compute the overdensities at $z = 0$, but this would require a further assumption that 
we observe the clusters just after their collapse. However, due to low $z$ values in our sample, our results do not change significantly, whether we compute the overdensities at $z = 0$ or at the observed 
redshifts, and we report our results using the latter method.

Figure 2 shows that $M_{1000}$ is strongly correlated with $\langle T \rangle$, the X-ray emission weighted cooling-flow corrected temperature. We fit the masses with a function 
\begin{equation}
M_{i} = n_{i}  \left({\langle T \rangle_{z = 0} \over {\rm 10 \ keV}}\right)^{\alpha_{i},} 
\end{equation}
where i denotes different values of overdensity. 

We are able to obtain interesting constraints for the above relation for radii up to $r_{1000}$. We exclude the low temperature data from the fits at larger radii, because the gas in the nearby ROSAT objects
is not detected to such low overdensities. 
At $r_{1000}$ we get an acceptable fit with $n_{1000} = 1.23^{+0.21}_{-0.18} \times \ 10^{15} \ 
h_{50}^{-1} \ M_{\odot} $,  $\alpha_{1000} = 1.79^{+0.14}_{-0.13}$, $\chi^{2} = 6.4$ for 6 d.o.f. The self-similar prediction $\alpha = 3/2$ differs by 3.7 $\sigma$. If we use the higher ASCA temperature for
NGC507 (see 2.4.2) the relation steepens by a few per cent, and the difference with the self-similar slope increases slightly. If we were to include the data of NGC5044 in the fit at $r_{1000}$ 
(see 2.4.3.), the results would not change significantly since that data point is consistent with the ones of HCG62 and NGC507, but the constraint would improve. Therefore, our choices of using the lower 
temperature for NGC507 and excluding NGC5044 data at $r_{1000}$ in all our fits are conservative.  Fixing the slope $\alpha_{1000}$ to 1.5 the best fit gives $\chi^{2} = 29.4$ for 7 d.o.f., thus
$\alpha_{1000}$ = 3/2 is ruled out at 99.99\% confidence. In the radial range  $r_{2000} \le r \le r_{1000}$, the values of $\alpha$ are consistent with a constant and significantly larger than 1.5 (see 
Table 2). However, if only the 6 hotter ASCA clusters are fitted, the slope at $r_{1000}$ is $\alpha_{1000} = 1.8 \pm 0.5$, consistent with 3/2 within the large errors.

\section{DISCUSSION}
\subsection{The slope of the $M - T$ relation}
Our main result is that the M-T scaling in the 1 - 10 keV range is inconsistent with the self-similar prediction. A possible explanation for the steeper slope of the M-T relation, compared to the self-similar
one, is preheating of the intracluster gas by supernova driven galactic winds before the clusters collapse, as proposed by e.g. David et al. (1991), Evrard \& Henry (1991),  Kaiser (1991), Loewenstein
\& Muskotzky (1996), to explain other X-ray data, such as the $L_{X}-T$ relation and cluster elemental abundances. If supernovae release a similar amount of energy per unit gas mass in hot and cool clusters, 
the coolest clusters would be affected more significantly and exhibit a stronger shift to higher temperatures in the M-T diagram (see Figure 2) than the hotter clusters. This will steepen the M-T 
relation. The cluster formation simulations by Metzler \& Evrard (1997), which include supernova heating, produce a slightly steeper slope (M$\propto T^{1.61}$) compared to the self similar slope 1.5 in 
their results of simulations with no winds. They show that if all the wind energy is thermalized and retained within a virial radius, the temperature at masses $\sim 10^{13} M_{\odot}$ may increase by 100\%, 
which would totally break the scaling at low temperatures. They note that in reality the effect should not be that dramatic since the extra energy is spent on work to lift the gas within the cluster potential
well. 

Our results are consistent with this explanation. Fitting only the ASCA data points (kT $>$ 4 keV) at $r_{1000}$ with $\alpha_{1000}$ fixed to 3/2 leads to an acceptable fit with 
$n_{1000} = 1.06^{+0.08}_{-0.09}$ and $\chi^{2} = 6.7$ for 5 d.o.f. (see Figure 2). 
HCG62, NGC507 and NGC5044 are then 50\%, 30\% and 25\% hotter than what the extrapolation of the self-similar fit would predict for their masses.
These amounts are reasonable in the supernova heating scheme, according to the above simulation (Metzler \& Evrard 1997).
 
Other work is also consistent with supernova heating scenario. Horner et al. (1999) examined the cluster M-T relation using several methods and found that spatially resolved temperature profile analysis for a
sample of clusters with kT $>$ 3 keV gives $\langle T \rangle^{3/2}$ scaling, whereas their isothermal analysis sample that includes some cooler groups with  kT $>$ 1 keV gives a steeper slope, consistent 
with ours. Hjorth et al. (1998) compare temperatures and gravitational lensing masses of clusters with kT $>$ 5 keV and obtain an M-T slope consistent with the self-similar one. The analysis of a sample of 
clusters with kT $>$ 4 keV (Neumann \& Arnaud 1999) shows that the self-similar slope of 3/2 that follows from their isothermal assumption, is consistent with their data. These observations, together with our
present results, are consistent with the hypothesis that at high temperatures the M-T scaling is self-similar, but breaks down at low temperatures ($\sim 1 $ keV). However, Horner et al. (1999) also compare 
temperatures with masses determined from velocity dispersions using the virial theorem. That dataset appears to be consistent with the self-similar scaling, even though their sample contains clusters with 
temperatures as low as $\sim$ 2 keV. The source of this disagreement is unclear (see 4.2. for more discussion on the virial masses). 

Also supportive of energy injections, the work of Ponman, Cannon \& Navarro (1999) shows that cool (T $<$ 4 keV) clusters observed with ROSAT and GINGA have entropies higher than achievable through 
gravitational collapse alone which they explain by pre-heating from strong galactic winds.

In the hydrostatic equilibrium scheme, since approximately $M_{tot} \propto T \times \beta$ at a given radius, for a given object with a certain total mass, a temperature rise due to preheating will be 
compensated by a shallower gas density profile. If heating is more prominent at small temperatures, one would then expect lower values of $\beta$ in cooler objects. This seems to be consistent with 
observations (e.g. Mohr \& Evrard 1997, Vikhlinin et al. 1999, Ponman et al. 1999). Our sample also exhibits this behaviour. The average values of $\beta$ with rms errors for the ROSAT and the ASCA subsample 
are $0.54 \pm 0.09$ and $0.70 \pm 0.06$, respectively.

\subsection{The normalization}
$M_{1000}$ values given by our best fit model with the exponent as a free parameter are 2.8 and 1.4 times smaller than corresponding values obtained by the Evrard et al. (1996) $\Omega = 1$ simulations 
at T = 1 keV and 10 keV, respectively. Note that both ASCA and ROSAT data, which have been obtained using independent temperature measurement methods and different instruments, give smaller values compared 
to the simulations. The above mentioned temperature profile analysis by Horner et al. (1999) gives a normalization 40\% lower than Evrard et al. (1996) at $r_{200}$, which is the same difference as we find 
between the normalization of our $\alpha \equiv 3/2$ fit and Evrard et al. (1996) at $r_{1000}$. The normalization of the isothermal sample results of Neumann \& Arnaud (1999) also is 30\% lower than the 
Evrard et al. (1996) values at $r = 0.3 \ r_{200}$. It is not inconceivable that the X-ray-measured masses are a factor of $\sim$2 lower than the true masses (in simulations), due, for example, to 
significant gas turbulence or magnetic fields (e.g. Loeb \& Mao 1994) invoked to explain the difference between the X-ray and lensing mass measurements. However, this seems increasingly unlikely as the 
cooling flow and temperature gradient effects appear to account for most of that disagreement (Allen 1998; Markevitch et al. 1999).  

The gravitational lensing mass analysis by Hjorth et al. (1998) gives a 12\% lower  $M - T$ normalization than Evrard et al. (1996), while Sadat, Blanchard \& Oukbir (1998) found gravitational lensing masses
in their sample to be 36\% below the Evrard et al. (1996) scaling. Lensing masses are highly uncertain at present (see e.g. Hjorth et al. 1998). Within the errors, both above results are in agreement with 
our and other X-ray results.      

On the other hand, the virial sample in Horner et al. (1999) gives a normalization consistent with Evrard et al. (1996). There are limitations for each of the three mass measurement methods (virial, X-ray, 
lensing). For example, the virial masses may be inflated by inclusion of background and foreground galaxies. The comparison is best done on the case-by-case basis, which is out of the scope of this paper.
The qualitative agreement of all other results suggests that it is the simulated values by Evrard et al. (1996) that may be incorrect. These simulations also produce too steep gas density profiles and too 
shallow temperature profiles compared to observations (e.g. Vikhlinin et al. 1999, Markevitch et al. 1998, respectively). Comparison of several independent cluster formation codes indicates that gas 
temperature and density profiles are areas where different simulations disagree significantly (Frenk et al. 1999). For example, simulations by Bryan \& Norman (1997) predict temperature profiles similar to 
those observed and should therefore produce an M-T relation closer to that observed.

\section{CONCLUSIONS}

We studied a sample of 9 relaxed galaxy clusters, galaxy groups and galaxies whose temperatures range from 1-10 keV and for which accurate temperature profiles are available. For the hotter subsample, the 
hydrostatic total mass profiles have been accurately determined up to $r_{500}$ using gas temperature profiles measured with ASCA. For the cooler subsample, the mass profiles are determined from spatially 
resolved spectroscopy of the ROSAT PSPC up to radii $r_{1500}$ - $r_{1000}$. The mass profiles of the cool subsample are consistent with the ``universal'' NFW model, as has earlier been found for the hotter 
subsample. We derived the mass-temperature relation at $r_{1000}$ over the 1-10 keV temperature range as

\begin{equation}
M_{1000} = 1.23 \pm 0.20 \ 10^{15} \ h_{50}^{-1} \ M_{\odot}
\left({{\langle T \rangle_{z = 0}} \over {\rm 10 \ keV}}\right)^{1.79 \pm 0.14}.
\end{equation}
The normalization is significantly smaller than that predicted by simulations of Evrard et al. (1996) with $\Omega = 1$. Our relation is significantly steeper compared to the self-similar one (slope of 3/2).
However, fitting only our ASCA data (kT $>$ 4 keV) at $r_{1000}$ with $\alpha_{1000}$ fixed to 3/2 leads to an acceptable fit 
\begin{equation}
M_{1000} = 1.06 \pm 0.09 \ 10^{15} \ h_{50}^{-1} \ M_{\odot}
\left({{\langle T \rangle_{z = 0}} \over {\rm 10 \ keV}}\right)^{1.5},
\end{equation}
with 2.6 $\sigma$ lower normalization than in the above mentioned simulations (Evrard et al. 1996).
Although we cannot exclude a single power law slope over the whole temperature range, this behaviour is consistent with a break in the self-similar scaling at low temperatures, as expected if preheating of 
the intracluster gas by supernova driven galactic winds has taken place. Most independent M-T relation observations are consistent with our results and with the preheating scenario. The gas density slopes in 
our low temperature sample are smaller than the ones in the hot subsample, consistent with this supernova heating scheme.
 
XMM and Chandra missions will be useful for extending the sample size. Their large effective area, excellent angular resolution and better energy resolution over a wide energy range will improve the accuracy 
of the spatially resolved spectroscopy manyfold compared to the presently available data.

\acknowledgments
JN thanks Harvard Smithsonian Center for Astrophysics for the hospitality. JN thanks the Smithsonian Institute for a Predoctoral Fellowship, and the Finnish Academy for a supplementary grant. WF and MM 
acknowledge support from NASA contract NAS8-39073. We thank the referee for useful comments.

\begin{deluxetable}{lcccccccccc}
\scriptsize
\tablecaption{The data \label{tab1}}
\tablewidth{0pt}
\tablehead{\colhead{name} & \colhead{$r_{2000}$}  & \colhead{$r_{1500}$}  & \colhead{$r_{1000}$} & \colhead{$r_{500}$} & \colhead{$M_{2000}$}           & \colhead{$M_{1500}$}           & \colhead{$M_{1000}$}           & \colhead{$M_{500}$}            & \colhead{$<T>$} & \colhead{z}\\
\colhead{}                & \colhead{Mpc}         & \colhead{Mpc}         & \colhead{Mpc}        & \colhead{Mpc}       & \colhead{$10^{14} M_{\odot}$}  & \colhead{$10^{14} M_{\odot}$}  & \colhead{$10^{14} M_{\odot}$}  & \colhead{$10^{14} M_{\odot}$}  & \colhead{keV}   & \colhead{}}
\startdata
A2029   & 0.87 & 1.02 & 1.23 & 1.68 &  4.8$^{+1.8}_{-1.7}$   & 5.7$^{+2.0}_{-2.2}$    & 6.6$^{+2.7}_{-2.5}$    & 8.5$^{+3.5}_{-3.6}$     & $9.10\pm1.0$ &  0.0767\\ 
A401    & 0.92 & 1.06 & 1.27 & 1.73 &  5.6$^{+1.5}_{-1.1}$   & 6.4$^{+1.6}_{-1.3}$    & 7.5$^{+1.7}_{-1.5}$    & 9.4$^{+2.4}_{-2.2}$     & $8.0\pm0.4$ &  0.0748\\ 
A2256   & 0.91 & 1.07 & 1.28 & 1.73 &  5.2$^{+1.1}_{-1.0}$   & 6.1$^{+1.3}_{-1.2}$    & 7.3$^{+1.5}_{-1.5}$    & 8.8$^{+1.8}_{-1.8}$     & $7.3\pm0.5$ &  0.058\\ 
A3571   & 0.91 & 1.05 & 1.26 & 1.68 &  5.0$^{+1.0}_{-1.0}$   & 5.6$^{+1.1}_{-1.1}$    & 6.5$^{+1.3}_{-1.3}$    & 7.8$^{+1.6}_{-2.2}$     & $6.9\pm0.2$ &  0.040\\ 
A2199   & 0.71 & 0.81 & 0.96 & 1.31 &  2.2$^{+0.5}_{-0.5}$   & 2.5$^{+0.5}_{-0.5}$    & 2.8$^{+0.6}_{-0.6}$    & 3.5$^{+0.7}_{-0.7}$     & $4.8\pm0.2$ &  0.0299\\ 
A496    & 0.73 & 0.83 & 0.99 & 1.33 &  2.5$^{+0.5}_{-0.5}$   & 2.8$^{+0.6}_{-0.6}$    & 3.1$^{+0.6}_{-0.6}$    & 3.7$^{+0.8}_{-0.8}$     & $4.7\pm0.2$ &  0.0331\\ 
HCG62   & 0.30 & 0.34 & 0.42 & 0.57 & 0.16$^{+0.05}_{-0.05}$ & 0.19$^{+0.06}_{-0.06}$ & 0.22$^{+0.07}_{-0.07}$ & 0.29$^{+0.09}_{-0.09}$\tablenotemark{*}  & $1.16\pm0.08$ &  0.0138\\ 
NGC5044 & 0.32 & 0.37 & 0.45 & 0.62 & 0.20$^{+0.04}_{-0.04}$ & 0.23$^{+0.05}_{-0.05}$ & 0.27$^{+0.05}_{-0.05}$\tablenotemark{*} & 0.36$^{+0.07}_{-0.07}$\tablenotemark{*} & $1.09\pm0.03$ &  0.0087\\ 
NGC507  & 0.31 & 0.35 & 0.44 & 0.68 & 0.18$^{+0.04}_{-0.04}$ & 0.20$^{+0.04}_{-0.04}$ & 0.25$^{+0.05}_{-0.05}$ & 0.49$^{+0.12}_{-0.10}$\tablenotemark{*}  & $1.09\pm0.03$ &  0.0162\\
\tablenotetext{} {Masses and radii are evaluated using $H_{0} = 50 \
km \ s^{-1} \ Mpc^{-1}$. The tabulated temperatures are the observed ones, not redshift-corrected. Errors are given at 90\% confidence.}
\tablenotetext{*} {Excluded from the fit}
\enddata
\end{deluxetable}

 \begin{deluxetable}{lcccccccc}
\small
\tablecaption{Fit results \tablenotemark{} \label{tab2}}
\tablewidth{0pt}
\tablehead{\colhead{} & \colhead{$r_{2000}$}          & \colhead{}              & \colhead{$r_{1500}$}          & \colhead{}              & \colhead{$r_{1000}$}          & \colhead{}              & \colhead{$r_{500}$}           & \colhead{}  \\
           \colhead{} & \colhead{$\alpha \equiv 1.5$} & \colhead{$\alpha$ free} & \colhead{$\alpha \equiv 1.5$} & \colhead{$\alpha$ free} & \colhead{$\alpha \equiv 1.5$} & \colhead{$\alpha$ free} & \colhead{$\alpha \equiv 1.5$} & \colhead{$\alpha$ free}}
\startdata
n                 & 0.65$^{+0.04}_{-0.05}$  & 0.93$\pm$0.14 & 0.73$\pm$0.05 & 1.05$^{+0.16}_{-0.15}$  & 0.89$^{+0.06}_{-0.07}$ & 1.23$^{+0.21}_{-0.18}$ & 1.28$\pm$0.12 & 1.55$^{+0.50}_{-0.43}$ \\ 
$\alpha$          &                         & 1.77$\pm$0.11 &               & 1.77$^{+0.12}_{-0.11}$  &                        & 1.79$^{+0.14}_{-0.13}$ &               & 1.84$^{+0.51}_{-0.52}$ \\ 
$\chi^{2}$/d.o.f. & 33.9/8                  & 7.6/7         & 34.5/8        & 7.4/7                   & 29.2/7                 & 6.4/6                  & 5.1/5         & 3.2/4                   \\ 
%\tablenotetext{} 
\enddata
\end{deluxetable}

\newpage

\begin{figure*}
\psfig{figure=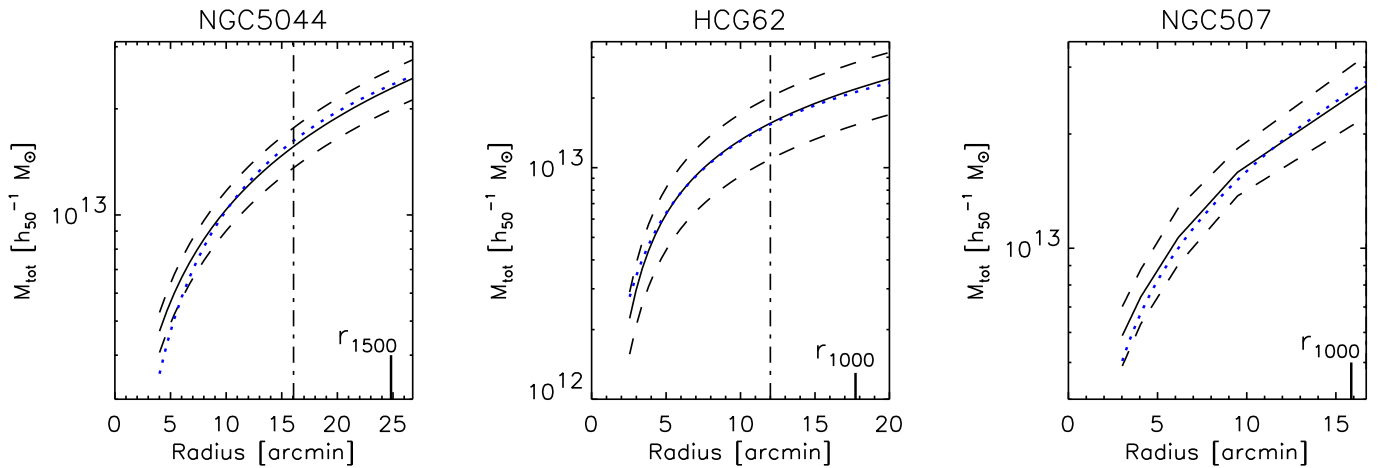,width=19cm,angle=0} 
\caption{The mass profiles for the ROSAT subsample (solid line) with 90\% confidence errors (dashed line). The profiles are plotted in the radial range beyond the cooling flow region and up to the maximum
radius of the significant surface brightness detection. The vertical dash-dot line shows the maximum radius of temperature measurement (in case of NGC507 it is equal to the maximum radius of brightness 
detection). The dotted line shows the total mass model with NFW profile for the dark matter component. \label{massprof_plot}}
\end{figure*}

\begin{figure*}
\psfig{figure=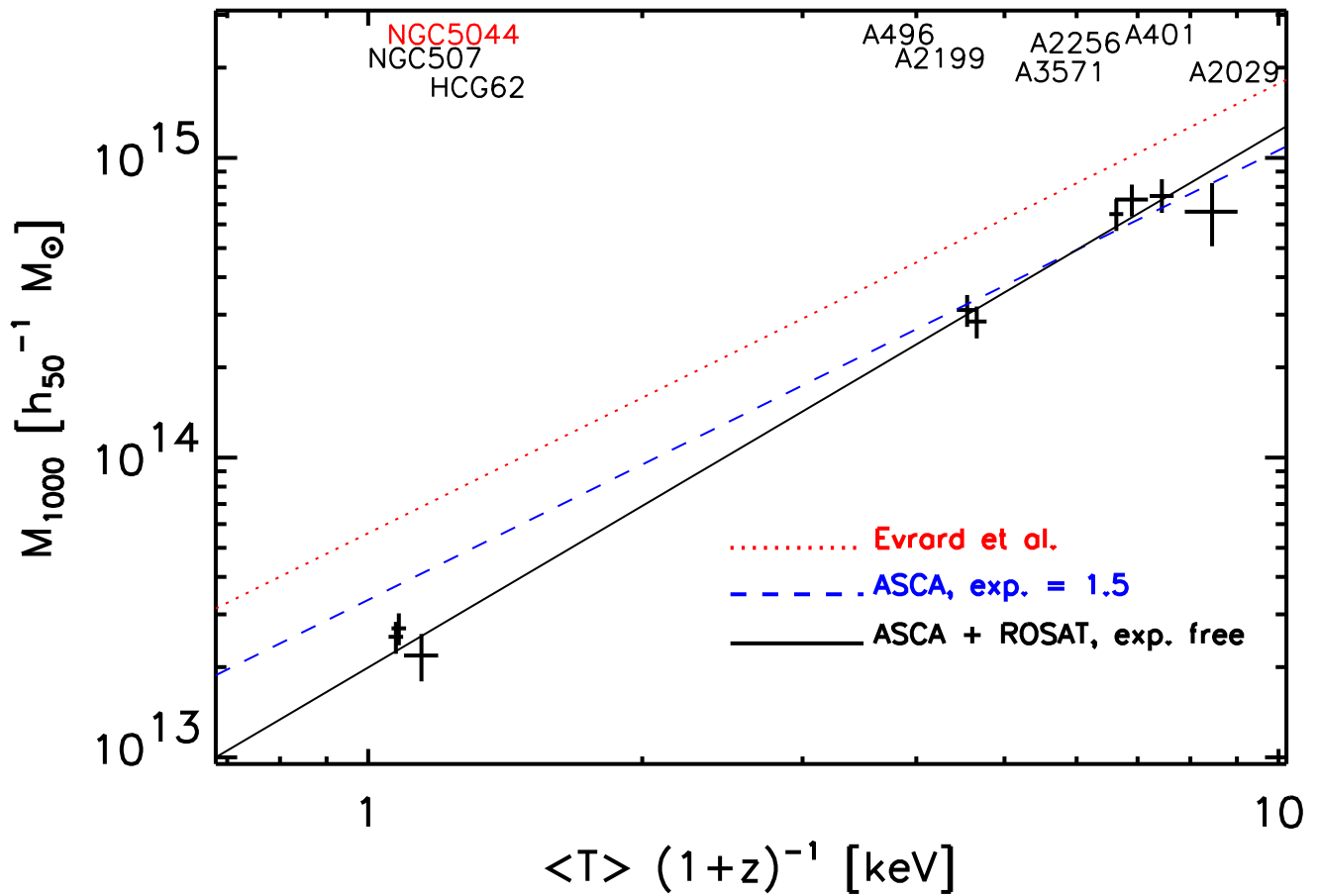,width=19cm,angle=0} 
\caption{Temperatures (scaled to z = 0) and $M_{1000}$ values with 1 $\sigma$ errors are plotted as crosses.  The best fit model to the ASCA data with the exponent fixed to 1.5 is shown as a dashed line. 
The best fit powerlaw to all data (excluding NGC5044, which is shown for illustration only) with exponent as a free parameter is shown as a solid line. The scaling law predicted by simulations (Evrard et
al. 1996) is shown as a dotted line. \label{mtrel_plot}}
\end{figure*}

\end{document}